\documentclass[12pt,aps,pra,eqsecnum,psfig,amssymb,subeqn,graphicx]{article}
\usepackage{amssymb}
 \newcommand{\eqref}[1]{(\ref{#1})}

\newcommand{\ds}{\displaystyle}
\newcommand{\be}{\begin{equation}}
\newcommand{\ee}{\end{equation}}
\newcommand{\ba}{\begin{array}}
\newcommand{\ea}{\end{array}}
\newcommand{\beqa}{\begin{eqnarray}}
\newcommand{\eeqa}{\end{eqnarray}}
\newcommand{\beqas}{\begin{eqnarray*}}
\newcommand{\eeqas}{\end{eqnarray*}}
\newcommand{\beqal}{\begin{lefteqnarray}}
\newcommand{\eeqal}{\end{lefteqnarray}}

\newcommand{\sdir}{\ensuremath{\rlap{\raisebox{.15ex}{$ \mskip 6.0 mu\scriptstyle+ $ }}\supset}\,\,}
\begin{document}
\title{Two oscillators quantum groups and associated deformed coherent states}

 \author{Nibaldo Alvarez--Moraga  \thanks{email address:
nibaldo.alvarez.m@exa.pucv.cl}\\  \small{D\'epartement de
Math\'ematiques et de Statistique, }
\\ \small{Universit\'e  de Montr\'eal, C.P. 6128,
Succ.~Centre-ville, Montr\'eal (Qu\'ebec), H3C 3J7, Canada} }

\maketitle

\begin{abstract}
Starting from a faithful five-dimensional matrix representation of
the group of two independent oscillators and applying the
$R$-matrix method we generate some classes of deformed
fermionic-bosonic quantum Hopf algebras. The corresponding Lie
deformed superalgebras of type I--II, obtained by duality, are
computed and a realization of generators of these deformed
superalgebras are given in terms of the usual fermionic and
bosonic  creation and annihilation operators associated to the
supersymmetric harmonic oscillator. Then, a generalized deformed
annihilator is construted and their eigenstates are computed
giving a new class  of deformed  coherent states.
\end{abstract}

%\pacs{11.30.-j, 03.65.Fd, 02.20.-a}

\baselineskip 0.74cm

\newpage

\setcounter{equation}{0}\section{Introduction} Quantum groups
\cite{kn:Ma95,kn:Dri86,kn:Reshe} are Hopf algebras equiped with a
quasitriangular structure. They are generalizations of Lie groups
and  algebras. Since their creation in the mid-eighteen, quantum
groups have attracted considerable attention in theoretical and
mathematical physics because the richness of their  algebraic
structure reflected in useful technical elements such as coproduct,
twisted product and counit, linked to many potential applications.
For example, in relation with these three properties it have been
established that quantum field are an example of
infinite-dimensional quantum groups \cite{kn:BrOr}.  In general, the
concept of deformed quantum Lie algebras has found various
applications in quantum optics, quantum field theory, quantum
statistical mechanics, supersymmetric quantum mechanics and some
purely mathematical problems. For instance, in the case of boson
quantum algebras, the special coproduct properties are useful to
characterize multi-particle Hamiltonians \cite{kn:TsoPaJa}.
Recently, deformed coherent and squeezed states have been associated
to quantum Heisenberg algebras \cite{kn:Nalvar4}. In the case of the
Poincar\'{e} quantum algebra, the coproduct have been brought to
bear the study the fusion of phonons \cite{kn:Cele2}. In the case of
the $su_q (2)$ algebra, it has been found that the $su_q (2)$
effective Hamiltonians reproduce accurately the physical properties
of the $su(2) \oplus h(2)$ models \cite{kn:BaCiHeRe}.

The goal of this article is firstly, to use the $R$--matrix method
\cite{kn:Ma95,kn:Dri86,kn:FaRe}  to generate some deformations of
the group of two independent oscillators and the corresponding
deformed quantum Lie algebras. More precisely, we consider a
faithful five-dimensional matrix representation of this group and
we apply the $R$--matrix method to compute the deformed quantum
groups  and quantum Lie superalgebras of  fermionic-bosonic type.
Next, we will find a  representation of these deformed
algebras and construct some classes of deformed coherent states.

We recall that, the $R$--matrix method has been used to generate
quadratic relations between the basic elements of a given group,
considered as generators of a bialgebra with coproduct derived
from the group law.  In this way, deformed bialgebras have been
defined whose external consistency is ensured if $R$ satisfies the
well-known quantum Yang-Baxter equation (QYBE). The associated
deformed quantum Lie algebras (or superalgebras) have been
obtained by defining a suitable duality operation.

This method has been applied, for instance, to the $ GL_q (2,
{\mathbb C})$
 and $ GL_{p,q} (2, {\mathbb C})$
matrix quantum groups (\cite{kn:Sudbe} and  \cite{kn:Dobrev},
respectively), the $GL(2,{\mathbb C})$ matrix group \cite{kn:Hieta},
the $GL(1 \| 1)$  supergroup \cite{FH97}, Heisenberg group
\cite{kn:HL94} and oscillator group \cite{HLR96}. In the case of the
Heisenberg group, the only deformed Lie algebras that can be
obtained are of the  bosonic type whereas in the cases of the $GL(2,
{\mathbb C})$ and oscillator groups these are of the fermionic and
bosonic type.

Considering the case of the group of two independent oscillators,
three types of deformed Lie algebras appear, i.e.,
fermionic-fermionic, fermionic-bosonic and bosonic-bosonic ones.
Although these three types of deformed algebras are interesting,
we pay special attention to the fermionic-bosonic type because
these algebras contain as sub-algebras the deformed
Heisenberg-Weyl Lie superalgebra which introduces new features
concerning the study of deformed coherent states.

This article is organized as follows. In section \ref{sec-dos}, we
give a description of the $R$-matrix method by applying it to a
five-dimensional faithful matrix representation of the group of two
independent oscillators.  In section \ref{sec-tres}, we compute the
deformed fermionic-bosonic bialgebras of the type I-II and the
corresponding deformed Lie superalgebras associated to this group.
In section \ref{sec-Fock},  we give some  realizations of this
deformed superalgebras in terms of the usual creation and
annihilation operators associated to the standard and supersymmetric
harmonic oscillators.  In section \ref{sec-cinco}, based on the
preceding realizations, we define the deformed supersymmetric and
generalized harmonic oscillator annihilators, and we compute their
eigenstates. We interpret these eigenstates as coherent states
associated to the two oscillator quantum group. Details of some
calculus are given in Appendices \ref{sec-appa} and \ref{sec-appb}.

\setcounter{equation}{0}\section{Two oscillator quantum groups}
\label{sec-dos} In this section we describe the R-matrix method
\cite{kn:FaRe} to construct certain classes of deformed quantum
algebras associated to the direct product of two oscillator groups.

The starting point is a five-dimensional faithful matrix
representation of the group of two independent oscillators. An
element is given by \be T =\pmatrix{ 1 & \alpha & \beta & 0  & 0
\cr
 0  &  \eta & \gamma & 0 & 0  \cr
 0  &  0 &  1 & 0 & 0   \cr
0 & 0 &  c & d &0 \cr
 0 & 0 & b & a &1 \cr},
\ee where the parameters $ \alpha, \beta, \gamma, \eta,  a, b, c,
d $ and the unity $1$ are considered as the generators of a
commutative algebra ${\cal A},$ the space of linear functions of
these generators, provided with a structure of Hopf algebra by
tensorial multiplication: \beqa \nonumber \Delta 1 &=& 1 \otimes
1, \\
\Delta \alpha &=& 1 \otimes \alpha + \alpha \otimes \eta, \qquad
\Delta \beta = 1 \otimes \beta + \beta \otimes 1 + \alpha \otimes \gamma, \nonumber \\
\nonumber \Delta \gamma &=& \eta \otimes \gamma + \gamma \otimes
1, \qquad  \Delta \eta = \eta \otimes \eta,  \\
\nonumber \Delta a &=& 1 \otimes a + a \otimes d,  \qquad  \Delta
b =1 \otimes b + b \otimes 1 + a \otimes c,
\\
\Delta c &=& d \otimes c + c \otimes 1,  \qquad
  \Delta d = d \otimes d. \label{copro}  \eeqa
According to this co-multiplication law, the generators of the
corresponding Lie algebra, in the representation space ${\cal A}$,
are given by \beqa X_1  &=& {\partial \over
\partial \alpha}, \qquad X_2 ={\partial \over
\partial \beta}, \\  X_3 &=& \alpha {\partial \over
\partial \beta} + \eta {\partial \over \partial \gamma}, \qquad
X_4 = \eta {\partial \over
\partial \eta} + \alpha {\partial \over \partial \alpha}, \\
{\tilde X}_1  &=& {\partial \over
\partial a}, \qquad {\tilde X}_2 ={\partial \over
\partial b}, \\ \qquad {\tilde X}_3 &=& a {\partial \over
\partial b} + d {\partial \over \partial c}, \qquad
{\tilde X}_4 = d {\partial \over
\partial d} + a {\partial \over \partial a}. \label{x123-diff}
\eeqa They verify the commutation relations of the Lie algebra $
ho  (4,{\mathbb R}) \oplus ho ({4,\mathbb R}),$ constructed as the
direct sum of the well-known Lie algebras associated with two
independent harmonic oscillators .  The non-zero commutation
relations of this Lie algebra are given by \beqa \label{he-we1}
[X_1,X_3]= X_2, \qquad [X_4, X_1]= - X_1, \qquad  [X_4, X_3]= X_3, \\
\biggl[ {\tilde X }_1, {\tilde X }_3 \biggr] = {\tilde X }_2,
\qquad [ {\tilde X}_4, {\tilde X}_1 ]= - {\tilde X }_1, \qquad
[{\tilde X }_4, {\tilde X }_3]= {\tilde X }_3. \label{he-we2}
\eeqa

The non-deformed quantum Lie algebra associated to the two
oscillator group corresponds to the dual space of ${\cal A}.$ The
action of their generators, $ A,B,C,H,{\tilde A}, {\tilde B},
{\tilde C}, {\tilde H}$ on the generating  elements  ${\cal P} =
\beta^{k} \eta^{l} \alpha^m \gamma^{n}  b^{r} a^s d^t c^{u}, \
k,l,m,n,r,s,t,u  \in {\mathbb Z}_{+} ,$  of ${\cal A}$, is given by
\cite{kn:Sudbe} \beqa (A, {\cal P}) &=& \biggl{( X_1 \cal P
\biggr)}_{\{0\}} = \delta_{k0
}  \, \delta_{m1} \,  \delta_{n0} \delta_{r0 } \, \delta_{s0} \,  \delta_{u0},\label{A-dual} \\
(B, {\cal P}) &=& \biggl{( X_2 \cal P \biggr)}_{\{0\}} = \delta_{k1
} \, \delta_{m0}  \, \delta_{n0}  \, \delta_{r0 }  \, \delta_{s0}  \, \delta_{u0}, \label{B-dual}\\
(C, {\cal P}) &=& \biggl{( X_3 \cal P \biggr)}_{\{0\}} = \delta_{k0
} \, \delta_{m0}  \, \delta_{n1}  \, \delta_{r0 }  \, \delta_{s0} \, \delta_{u0},\label{C-dual} \\
(H, {\cal P}) &=& \biggl{( X_4 \cal P \biggr)}_{\{0\}} = l \,
\delta_{k0
} \, \delta_{m0} \, \delta_{n0} \, \delta_{r0 } \, \delta_{s0}  \, \delta_{u0},\label{D-dual} \\
({\tilde A}, {\cal P}) &=& \biggl{( {\tilde X}_1 \cal P
\biggr)}_{\{0\}} = \delta_{k0
}  \, \delta_{m0}  \, \delta_{n0} \, \delta_{r0 } \, \delta_{s1}  \, \delta_{u0},\label{tildeA-dual} \\
({\tilde B}, {\cal P}) &=& \biggl {( {\tilde X}_2 \cal P
\biggr)}_{\{0\}} = \delta_{k0
} \, \delta_{m0}  \, \delta_{n0} \, \delta_{r1 } \, \delta_{s0} \, \delta_{u0}, \label{tildeB-dual}\\
({\tilde C}, {\cal P}) &=& \biggl{( {\tilde X}_3 \cal P
\biggr)}_{\{0\}} = \delta_{k0
} \, \delta_{m0}  \, \delta_{n0} \, \delta_{r0 } \, \delta_{s0} \,  \delta_{u1}, \label{tildeC-dual}\\
({\tilde H}, {\cal P}) &=& \biggl{( {\tilde X}_4 \cal P
\biggr)}_{\{0\}} =t \, \delta_{k0 } \,  \delta_{m0} \, \delta_{n0}
\, \delta_{r0 } \, \delta_{s0}  \, \delta_{u0},
\label{tildeD-dual}\eeqa where $\{0\}\equiv
{\{\alpha=\beta=\gamma=0, \eta=1; a=b=c=0, d=1\}}.$ The action of
the product of two of these generators, on an arbitrary element
$\cal P,$ is computed with the help of the homomorphism property of
the co-multiplication. That is, for generic generators $V$ and $W,$
we get \be (V W, {\cal P} ) = (V \otimes W, \Delta {\cal P})=
\sum_{(c)}(V \otimes W, {\cal P}^{(c)}\otimes {\cal P}^{(c)})=
\sum_{(c)} (V, {\cal P}^{(c)} ) (W, {\cal P}^{(c)} ),
\label{ho-pro-product}\ee where, in the usual notation, $ {\cal
P}^{(c)}$ represents the generic elements of $\cal A$ generated on
the co-multiplication action. Then, the commutator of this
generators is computed from \be ( [V , W], {\cal P} ) = ( V \otimes
W - W \otimes V, \Delta {\cal P}).\label{com-rel-coproduct} \ee

Following this procedure it is easy to show that the  generators of
the dual space of $\cal A$  satisfy the same non-zero commutation
relations as (\ref{he-we1}-\ref{he-we2}),
i.e., \beqa [A, C]= B, \qquad [H, A]= - A, \qquad [H, C]= C, \\
\biggl[ {\tilde A }, {\tilde C } \biggr] = {\tilde B }, \qquad [
{\tilde H}, {\tilde A} ]= - {\tilde A }, \qquad [{\tilde H },
{\tilde C }]= {\tilde C}. \eeqa

\subsection{The R-matrix method}

To obtain the possible associated deformed Lie algebras
(superalgebras), we apply the well-known $R$--matrix
method\cite{kn:Dri86}. Firstly, we deform the basic algebra ${\cal
A}$ with help  of a matrix $R$ which satisfies the quantum
Yang-Baxter equation(QYBE). Then the corresponding deformed quantum
algebras (superalgebras) are obtained by duality, according to the
relations (\ref{A-dual}--\ref{com-rel-coproduct}) (in the case of a
superalgebra we must replace in (\ref{com-rel-coproduct}) the
commutator by an anti-commutator and the sign $-$ by a sign $+$ when
both $V$ and $W$ represents generators of the type fermionic).

Let us introduce a $25 \times 25$ matrix $R$ defined by the
equation \be \label{reldef} R T_1 T_2 = T_2 T_1 R, \ee where \be
T_1 = T \otimes I, \qquad T_2 = I \otimes T, \ee with $I$ the $5
\times 5$ identity matrix, and satisfying the QYBE \be
\label{qybe} R_{12} R_{13}R_{23} = R_{23} R_{13}R_{12},\ee where
\be R_{12} = R \otimes I,  \qquad R_{23} = I \otimes R \ee and
$R_{13}$ have a similar expression. Writing $R$ in the form of
block  matrices formed by the $5 \times 5$ dimensional
sub-matrices $R^{ij}, \, i,j =1,2\ldots,5 $  and denoting their
entries  by ${(R^{ij})}_{kl} = r^{ij}_{kl}, \, k,l=1,2,\ldots,5,$
and the entries of $T$ by $t_{ij},$ we can write explicitly  the
deformed law \eqref{reldef} as \be \label{rel-def} r^{ij}_{kl} \,
t_{jm} \, t_{ls} \, = \, t_{kl} \, t_{ij} \, r^{jm}_{ls}, \ee with
summation over repeated indices. In the same way, the QYBE
\eqref{qybe} writes \be r^{ij}_{kl} \, r^{lm}_{np} \, r^{jq}_{sn}
\, =  \, r^{kl}_{sn} \, r^{ij}_{np} \, r^{jq}_{lm}. \ee
\subsection{Commutative bialgebra}
In the particular case where $R$ is the identity matrix,
$r^{ij}_{kl} = \delta^{ij} \delta_{kl},$ equation  \eqref{rel-def}
becomes \be t_{im} t_{ks} = t_{ks} t_{im}, \quad \forall \, i,m,k,s
= 1, 2, \ldots, 5, \ee i.e., we regain the commutative algebra
${\cal A}.$
\subsection{Non-commutative bialgebra}
Other diagonal $R$--matrix compatible with the coproduct
\eqref{copro} for which we get a non-commutative (non deformed)
bialgebra is given by the fermionic--bosonic type \linebreak
$R$--matrix, $R^{ij} = \delta^{ij} {\tilde I} $ where ${\tilde I} =
I$  when $ (i,j) \ne (2,2) $ and ${\tilde I} = {\rm diag} \,
(1,-1,1,1,1)$ when $ (i,j)  = (2,2).$ In this case \eqref{rel-def}
implies  \be \label{com-fer}
 \{\alpha , \eta\}=0, \qquad \{\gamma ,
\eta \}=0,   \qquad \alpha^2 =0, \qquad \gamma^2 = 0, \ee \be
[\alpha , \beta ]=0, \qquad [\alpha , \gamma ]=0, \qquad [\beta,
\gamma ]= 0 \qquad [\beta, \eta ]= 0 \ee \be [a, b]=0, \quad
[a,c]=0, \quad  [a,d]=0, \quad [b,c]=0, \quad [b,d]=0, \quad
[c,d]=0,  \label{com-bos}\ee
 and all the commutators between the Greek generators and the Roman generators being equal to zero.
Here  $\{,\}$ denotes the anti-commutator.

In this last case, the non-zero super-commutation relations
satisfies by the dual generators are given by  \beqa  \{A, C\}= B, \qquad [H, A]= - A, \qquad  [H, C]= C, \\
\ds  [{\tilde A} , {\tilde C} ] = {\tilde B }, \qquad [ {\tilde
H}, {\tilde A} ]= - {\tilde A }, \qquad [{\tilde D }, {\tilde C
}]= {\tilde C}, \eeqa with \be A^2=0, \qquad C^2=0. \ee These are
the commutation relations of  a  Lie superalgebra isomorphic to
 $ho (2 / 2, {\mathbb R} ) \oplus ho (4, {\mathbb R}).$

\setcounter{equation}{0}\section{Deformed fermionic-bosonic quantum
superalgebras} \label{sec-tres} We are interested to deform the
superalgebra $ho (2 / 2, {\mathbb R} ) \oplus ho (4, {\mathbb R}).$
To do it, firstly we find a set of deformed $R$ matrices,
continuously connected to the non deformed fermionic-bosonic type
$R$-matrix and solving \eqref{reldef}, then we demand  that they
verify the QYBE \eqref{qybe}. In Appendix \ref{sec-appa}, we get a
set of this type of deformed matrices and the  relations of
consistency between their entries. We notice that there are several
possible choices for the remaining entries of these $R$ matrices so
that they verify the QYBE. These different choices determinate the
different types of deformation of the fermionic-bosonic bialgebra
(\ref{com-fer}-\ref{com-bos}). Indeed, they enter in the
classification given in \cite{HLR96}, but with a highest number of
deformation parameters, and we get the type I--I, I--II, I--III
II--I II-II and II--III fermionic--bosonic bialgebras. In the
following, we give some examples of deformed bialgebras of  the I-II
fermionic--bosonic type and construct the corresponding quantum
superalgebras.

\subsection{Type I-II fermionic-bosonic bialgebras}
The type I-II fermionic-bosonic quantum groups are obtained by
setting $ r^{12}_{22} = r^{22}_{23}=0$ and
$r^{54}_{44}=r^{44}_{43}=0$ in \eqref{fer} and \eqref{bos},
respectively. Again, there are several possibilities to choose the
R-matrix verifying the QYBE \eqref{qybe}:

\subsubsection{Direct sum  bialgebra structure}
For instance, the choice
($\epsilon= \pm 1$)  \beqa r^{12}_{12} = z, \qquad   r^{23}_{23} = x, \\
r^{12}_{23}= r^{23}_{12} =  \epsilon \sqrt{z x}, \qquad r^{13}_{22}=
r^{13}_{11}
- \epsilon \sqrt{z x}, \qquad r^{22}_{13}= r^{11}_{13} - \epsilon \sqrt{z x}, \\
r^{43}_{14} =r^{53}_{15} =w, \qquad r^{53}_{43}= - r^{43}_{53} =
\rho, \qquad r^{44}_{13}= r^{55}_{13}= - w + r^{11}_{13}, \\
r^{53}_{44}=  - r^{44}_{53} =  q,  \qquad  r^{54}_{53}= = -
r^{53}_{54}=  \tau, \qquad r^{53}_{55}= - r^{55}_{53}= p + q, \qquad
\\ r^{13}_{33}=  r^{13}_{44} = r^{13}_{55} = r^{13}_{11}, \qquad
 r^{33}_{13}=  r^{11}_{13}, \qquad  r^{53}_{13}= - r^{13}_{53}, \eeqa
with arbitrary real parameters $x,z,p,q,w$ and coefficients $
r^{11}_{13}, r^{13}_{11}, r^{13}_{13}, r^{13}_{53}, $ produces the
deformed quantum algebra for which the non zero commutation
relations are given by \beqa  \label{part-fer} \{ \alpha , \eta \} =
0, \qquad \{ \gamma, \eta \} = 0, \qquad \alpha^2= {1\over 2} z (1 -
\eta^2 ), \qquad \gamma^2=
{1\over 2} x (1 -\eta^2) \\
\ds [\alpha , \beta] =  z  \gamma \eta, \qquad [\gamma ,\beta] = x
\alpha \eta \eeqa and \be \label{part-bos} [a,b]= p   a + \tau
(1-d), \qquad [b,c]= -q   c  - \rho (1-d), \ee i.e., the direct sum
of the deformed type I fermionic and type II  bosonic quantum groups
\cite{HLR96}.

The same deformation relations (\ref{part-fer}-\ref{part-bos}), but
with $q = - p,$ is obtained if
we take \beqa r^{12}_{12} = z, \qquad   r^{23}_{23} = x, \\
r^{12}_{23}= - r^{23}_{12} =  \epsilon \sqrt{z x}, \qquad
r^{13}_{22}= -
r^{22}_{13}= - r^{11}_{13} + \epsilon \sqrt{z x},   \\
r^{13}_{11}= - r^{11}_{13},
 \qquad  r^{13}_{13}= {z x \over 2} - {\left(r^{11}_{13} - \epsilon \sqrt{z x} \right)}^2 , \\\qquad
 r^{53}_{43}= - r^{43}_{53} = \rho, \qquad  r^{53}_{44}=  - r^{44}_{53} =  - p,  \qquad  r^{54}_{53} = -
r^{53}_{54}=  \tau, \qquad
\\ r^{13}_{33}=  r^{13}_{44} = r^{13}_{55} = - r^{44}_{13}= - r^{55}_{13}= - r^{33}_{13}= -  r^{11}_{13}  + 2 \epsilon   \sqrt{z x}, \qquad
 r^{53}_{13}= - r^{13}_{53}, \eeqa with  arbitrary real parameters
$x,z,p$ and coefficients $   r^{11}_{13}, r^{13}_{53}. $

\subsubsection{Non-direct sum bialgebra structure}
Other class of deformed quantum groups that do not show a direct sum
structure is given by the relations (\ref{part-fer}-\ref{part-bos}),
by taking  $q = -p,$ $\rho=\tau=0$ in addition to the commutation
relations \be \label{non-com1} [\beta ,a ] = \theta a, \qquad [\beta
, c] = - \theta c , \ee where $\theta $ is an additional arbitrary
parameter. Here the non zero matrix elements of
$R$ are given by \beqa r^{12}_{12} = z, \qquad r^{23}_{23} = x, \\
r^{12}_{23}= r^{23}_{12} =  \epsilon \sqrt{z x}, \qquad r^{22}_{13}=
r^{11}_{13} - \epsilon \sqrt{z x}, \qquad r^{13}_{22}= r^{13}_{11} -
\epsilon \sqrt{z x}, \\ r^{44}_{13}= r^{11}_{13} - \theta, \qquad
r^{13}_{44}= r^{13}_{11} + \theta,
\\ r^{13}_{33}= r^{13}_{55} =  r^{13}_{11}, \qquad r^{33}_{13}= r^{55}_{13}=  r^{11}_{13}, \\   r^{53}_{44}=  - r^{44}_{53} =  - p,
 \eeqa with arbitrary real parameters
$x,z,p, \theta $ and coefficients $   r^{11}_{13}, r^{13}_{11},
r^{13}_{13}, r^{53}_{53}. $

A more general quantum group  can be reached if we take \beqa
r^{12}_{12}=z, \qquad  r^{23}_{23}=x,
\\ r^{12}_{23}= -  r^{23}_{12} =
\epsilon \sqrt{z x}, \qquad r^{13}_{13} = { x z \over
2} -  {\left( r^{11}_{13} - \epsilon \sqrt{z x} \right)}^2, \\
r^{13}_{22} = - r^{22}_{13} = \epsilon \sqrt{z x} - r^{11}_{13},
\qquad r^{13}_{33}= r^{13}_{55} = - r^{33}_{13} = - r^{55}_{13}= 2
\epsilon \sqrt{z x} - r^{11}_{13}, \\ r^{13}_{11} =
 - r^{11}_{13},   \qquad r^{44}_{13} =
 - r^{13}_{44},  \qquad   r^{53}_{13} =
 - r^{13}_{53},
  \\  r^{53}_{43} =
 - r^{43}_{53}= \rho, \qquad  r^{54}_{53} =
 - r^{53}_{54}= \tau, \qquad r^{53}_{44}= - r^{44}_{53}= q,  \\ r^{13}_{44} = \sigma - r^{11}_{13} + 2 \epsilon \sqrt{ z x},\qquad
r^{13}_{43} = - r^{43}_{13} = {\rho  \sigma \over q}, \qquad
r^{13}_{54} = - r^{54}_{13} = - {\tau \sigma \over p},
\\r^{13}_{51}= r^{23}_{52}= r^{33}_{53}= r^{43}_{54}= -
r^{51}_{13}= - r^{52}_{23}=- r^{53}_{33}= - r^{54}_{43} = p - q,
\eeqa for arbitrary real parameters $x,z, p, q, \rho, \tau, \sigma $
and coefficients $r^{11}_{13}, r^{13}_{53}.$ In this case, the
corresponding deformed quantum group is given by
(\ref{part-fer}-\ref{part-bos}), in addition to the relations
 \be \label{non-com2} [\beta ,a ] =  {\tau \sigma \over p} (d-1) +  \sigma a, \qquad
 [\beta , c] ={\rho  \sigma \over q} (d-1)  - \sigma c.
\ee Let us notice that this results represent new classes of
deformed bialgebras having the structure which can not be
constructed  by considering only the direct sum between the
corresponding  type I fermionic and a type II bosonic oscillator
bialgebras.

\subsection{Deformed fermionic-bosonic quantum Lie superalgebras}
\label{sec-cuatro} In general, according to the coproduct law
\eqref{copro} and the duality relations
(\ref{A-dual}--\ref{tildeD-dual}), we can show that the type I-II
fermionic-bosonic deformed quantum Lie superalgebras have the
structure of a direct sum of the type I fermionic and type II
bosonic deformed quantum oscillator algebras obtained in
\cite{HLR96}. This is true even if the deformed bialgebras do not
have the structure of a direct sum,  as in
(\ref{part-fer}-\ref{part-bos},\ref{non-com1}) or in
(\ref{part-fer}-\ref{part-bos},\ref{non-com2}). We remark that this
last fact should play an important role at the moment to compute the
basic coproduct relations to the dual bialgebra level.

Indeed, from \eqref{ho-pro-product}, writing ${\cal P} = {\cal F}
{\cal B},$ where ${\cal F}= \beta^{k} \eta^{l} \alpha^m \gamma^{n},
\ k,l, \in {\mathbb Z}_+ , m,n=0,1 $ and ${\cal B}= b^{r} a^s d^t
c^{u},\ r,s,t,u \in {\mathbb Z}_+ ,$ using the homomorphism
properties of the coproduct \eqref{copro} and considering its
special structure as well as the structure of the bialgebras given
in the above subsection, for generic generators $V$ and $W,$ we get
\beqa \nonumber (V W, {\cal F} {\cal B}) &=& (V \otimes W, \Delta
({\cal F}
{\cal B} )) = (V \otimes W, (\Delta {\cal F})(\Delta {\cal B})) \\
\nonumber &=& \sum_{(c), (c^\prime)}(V \otimes W, ({\cal
F}_1^{(c)}\otimes {\cal F}_2^{(c)} ) \ ( {\cal
B}_1^{(c^\prime)}\otimes {\cal B}_2^{(c^\prime)})) \\ &=&  \sum_{(c)
\ (c^\prime)} (V, {\cal F}_1^{(c)} {\cal B}_1^{(c^\prime)}) \ (W,
{\cal F}_2^{(c)} {\cal B}_2^{(c^\prime)}). \label{fer-bos-part}\eeqa

In particular, if  $V,W  \in \{ A, B, C, H\},$  according  to the
duality relations (\ref{A-dual}--\ref{tildeD-dual}), we deduce \beqa
\nonumber (V W, {\cal F} {\cal B}) &=& \sum_{(c) \ (c^\prime)} \
  (V, {\cal F}_1^{(c)}) \ {({\cal B}_1^{(c^\prime)})}_{\{0\}} \
(W, {\cal F}_2^{(c)}) \ {({\cal B}_2^{(c^\prime)})}_{\{0\}}
\\   &=&  (1,{\cal B}) \ \sum_{(c)}  (V, {\cal
F}_1^{(c)}) \ (W, {\cal F}_2^{(c)})=  (VW, {\cal F})\ (1,{\cal B}).
\eeqa  From this result, we have for example  \be (\{V, W\}, {\cal
F} {\cal B}) = (\{V,W\}, {\cal F}) \ (1,{\cal B}), \label{com-F} \ee
i.e, only the fermionic  type part of the quantum bialgebra is
essential to compute the anti-commutator (commutator) for this class
of generators .

On the other hand, if  $V,W \in \{{\tilde A}, {\tilde B}, {\tilde
C}, {\tilde H}\},$ in the same way we can show that \be ([V, W],
{\cal F} {\cal B}) =(1, {\cal F}) \ ([V, W],{\cal B}), \label{ess-B}
\ee i.e, only the bosonic type part of the quantum bialgebra is
essential to compute the commutator for this class of generators.

Now, if  $ V \in \{ A, B, C, H\} $ and $W  \in \{{\tilde A}, {\tilde
B}, {\tilde C}, {\tilde H}\}, $ then from \eqref{fer-bos-part},
according to the  pairings  (\ref{A-dual}--\ref{tildeD-dual}), we
get \beqa \nonumber ([V ,W], {\cal F} {\cal B}) &=&
 \sum_{(c)
\ (c^\prime)} \biggl[   (V, {\cal F}_1^{(c)} ) \ {({\cal
B}_1^{(c^\prime)})}_{\{0 \}} \  {({\cal F}_2^{(c)})}_{\{ 0 \}} \ (W,
{\cal B}_2^{(c^\prime)}) \\ &-& {({\cal F}_1^{(c)})}_{\{0 \}} \ (W,
{\cal B}_1^{(c^\prime )} ) \   (V, {\cal F}_2^{(c)}) \ {({\cal
B}_2^{(c^\prime )})}_{\{ 0 \}} \biggr] . \label{dual-com-fer-bos}
\eeqa To know the explicit form of  \eqref{dual-com-fer-bos}, let us
write \cite{HLR96} \beqa \nonumber \Delta (\beta^k \eta^l) &=&
\Gamma^{k,i,j}_{v,w; v^\prime w^\prime} \ \beta^v \eta^{w+l}
\alpha^i \gamma^j \otimes \beta^{v^\prime} \eta^{w^\prime + l}
\alpha^i \gamma^j \\ &+& \Delta^{k,i,j}_{v,w; v^\prime w^\prime} \
\beta^v \eta^{w+l} \alpha^i \gamma^j \otimes \beta^{v^\prime}
\eta^{w^\prime + l} \alpha^{i+1} \gamma^{j+1},\label{beta-eta} \eeqa
\beqa \nonumber \Delta (\alpha^m \gamma^n) &=& \delta_{m 0}
\delta_{n 0} \ (1 \otimes 1) + \delta_{m 0} \delta_{n 1} \ (\eta
\otimes \gamma + \gamma \otimes 1) + \delta_{m 1} \delta_{n 0} \ (1
\otimes \alpha + \alpha \otimes \eta)
\\ &+&  \delta_{m 1} \delta_{n 1} \ ( \eta \otimes \alpha \gamma +
\alpha \gamma  \otimes \eta - \eta \alpha \otimes \eta \gamma  +
\gamma  \otimes \alpha),\label{alpha-gamma}  \eeqa and \be \Delta (
b^r a^s d^t c^u ) = \Lambda^r_{{\tilde k}, {\tilde l}, {\tilde m};
k^\prime, l^\prime, m^\prime, j^\prime} \ { s \choose s^\prime} { u
\choose u^\prime} \ b^{\tilde k} a^{{\tilde l} + s^\prime}
d^{{\tilde m} + t + u^\prime} c^{u - u^\prime} \otimes b^{ k^\prime}
a^{l^\prime + s - s^\prime} d^{m^\prime + t + s^\prime} c^{j^\prime
+ u^\prime}, \label{abcd}\ee with summation over repeated indices
and taking $i+1,j+1$ mod $2$ in \eqref{beta-eta}. The coefficients
$\Gamma^{k,i,j}_{v,w; v^\prime w^\prime},\ \Delta^{k,i,j}_{v,w;
v^\prime w^\prime}$ and $\Lambda^r_{{\tilde k}, {\tilde l}, {\tilde
m}; k^\prime, l^\prime, m^\prime j^\prime}$ satisfy certain  useful
recurrence relations that allow us to reduce efficaciously the
calculations (see reference \cite{HLR96}).

Combining \eqref{beta-eta} with \eqref{alpha-gamma}  to construct
$\Delta {\cal F},$ and interpreting both  this result  and
\eqref{abcd} according to \eqref{dual-com-fer-bos}, we get \beqa
\nonumber ([V ,W], {\cal F} {\cal B}) &=& \biggl\{ \nonumber
\delta_{m0} \delta_{n0} \ \biggl[ \Gamma^{k,00}_{vw;0 w^\prime} \
(V, \beta^v \eta^{w + l}) + \Delta^{k,11}_{vw;0 w^\prime} (V,
\beta^v \eta^{w + l}) \biggr]  \\
\nonumber &+& \delta_{m0} \delta_{n1} \ \biggl[ \Gamma^{k,00}_{vw;0
w^\prime} \ (V, \beta^v \eta^{w + l} \gamma ) + {1 \over 2}
\Delta^{k,11}_{vw;0 w^\prime}  (V, \beta^v \eta^{w + l} (1 - \eta^2
) \alpha ) \biggr] \\ \nonumber &+& \delta_{m1} \delta_{n0} \
\biggl[ \Gamma^{k,00}_{vw;0 w^\prime} \ (V, \beta^v \eta^{w + l}
\alpha ) + {1 \over 2} \Delta^{k,11}_{vw;0 w^\prime} (V, \beta^v
\eta^{w + l} (1 - \eta^2 ) \gamma ) \biggr]
\\ \nonumber &+&
\delta_{m1} \delta_{n1} \ \biggl[ \Gamma^{k,00}_{vw;0 w^\prime} \
(V, \beta^v \eta^{w + l} \alpha \gamma ) + {1 \over 4}
\Delta^{k,11}_{vw;0 w^\prime} (V, \beta^v \eta^{w + l} {(1 - \eta^2
)}^2) \biggr] \biggr\} \\
\nonumber &\times& \Lambda^r_{0,0,{\tilde m} ; k^\prime, l^\prime,
m^\prime, j^\prime} \ ( W, b^{ k^\prime} a^{l^\prime + s}
d^{m^\prime + t} c^{j^\prime + u}) \\ \nonumber &-&
 \Lambda^r_{{\tilde
k},{\tilde l}, {\tilde m}; 0, l^\prime, m^\prime, 0}\  { s \choose
l^\prime + s } \ (W, b^{\tilde k} a^{{\tilde l} + l^\prime + s}
d^{{\tilde m} + t} c^{u}) \\ \nonumber &\times&  \biggl\{
\delta_{m0} \delta_{n0}  \biggl[ \Gamma^{k,00}_{0w; v^\prime
w^\prime} \ (V, \beta^{v^\prime} \eta^{w^\prime + l}) +
\Delta^{k,00}_{0w; v^\prime w^\prime} \ (V, \beta^{v^\prime}
\eta^{w^\prime + l}) \biggr]
\\ \nonumber &+& \delta_{m0} \delta_{n1}  \biggl[ \Gamma^{k,00}_{0w; v^\prime w^\prime} \ (V, \beta^{v^\prime}
\eta^{w^\prime + l} \gamma) + {1 \over 2} \Delta^{k,00}_{0w;
v^\prime w^\prime} \ (V, \beta^{v^\prime} \eta^{w^\prime + l} (1 -
\eta^2) \alpha) \biggr]
\\ \nonumber &+& \delta_{m1} \delta_{n0}  \biggl[ \Gamma^{k,00}_{0w; v^\prime w^\prime} \ (V, \beta^{v^\prime}
\eta^{w^\prime + l} \alpha ) + {1 \over 2} \Delta^{k,00}_{0w;
v^\prime w^\prime} \ (V, \beta^{v^\prime} \eta^{w^\prime + l} (1 -
\eta^2) \gamma) \biggr] \\  &+& \delta_{m1} \delta_{n1} \biggl[
\Gamma^{k,00}_{0w; v^\prime w^\prime} \ (V, \beta^{v^\prime}
\eta^{w^\prime + l} \alpha \gamma ) + {1 \over 4} \Delta^{k,00}_{0w;
v^\prime w^\prime} \ (V, \beta^{v^\prime} \eta^{w^\prime + l} {(1 -
\eta^2)}^2 ) \biggr] \biggr\}, \qquad  \eeqa where summation over
$w$ and $w^\prime$ and over repeated indices is  supposed. We notice
that this last expression can be yet  simplified using  the pairings
(\ref{A-dual}--\ref{D-dual}), we get \beqa \nonumber ([V ,W], {\cal
F} {\cal B}) &=& \biggl\{ \nonumber \delta_{m0} \delta_{n0} \
\biggl[ \Gamma^{k,00}_{vw;0 w^\prime} \ (V, \beta^v \eta^{w + l}) +
\Delta^{k,11}_{vw;0 w^\prime} (V,
\beta^v \eta^{w + l}) \biggr]  \\
\nonumber &+& \delta_{m0} \delta_{n1} \ \Gamma^{k,00}_{vw;0
w^\prime} \ (V, \beta^v \eta^{w + l} \gamma ) +  \delta_{m1}
\delta_{n0} \ \Gamma^{k,00}_{vw;0 w^\prime} \ (V, \beta^v
\eta^{w + l} \alpha ) \biggr\} \\
\nonumber &\times& \Lambda^r_{0,0,{\tilde m} ; k^\prime, l^\prime,
m^\prime, j^\prime} \ ( W, b^{ k^\prime} a^{l^\prime + s}
d^{m^\prime + t} c^{j^\prime + u}) \\ \nonumber &-&
 \Lambda^r_{{\tilde
k},{\tilde l}, {\tilde m}; 0, l^\prime, m^\prime, 0}\  { s \choose
l^\prime + s } \ (W, b^{\tilde k} a^{{\tilde l} + l^\prime + s}
d^{{\tilde m} + t} c^{u}) \\ \nonumber &\times&  \biggl\{
\delta_{m0} \delta_{n0}  \biggl[ \Gamma^{k,00}_{0w; v^\prime
w^\prime} \ (V, \beta^{v^\prime} \eta^{w^\prime + l}) +
\Delta^{k,00}_{0w; v^\prime w^\prime} \ (V, \beta^{v^\prime}
\eta^{w^\prime + l}) \biggr]
\\  &+& \delta_{m0} \delta_{n1}  \ \Gamma^{k,00}_{0w; v^\prime w^\prime} \ (V, \beta^{v^\prime}
\eta^{w^\prime + l} \gamma) +  \delta_{m1} \delta_{n0}
\Gamma^{k,00}_{0w; v^\prime w^\prime} \ (V, \beta^{v^\prime}
\eta^{w^\prime + l} \alpha ) \biggr\}, \qquad \label{V-com-W}\eeqa
where summation over $w$ and $w^\prime$ and over repeated indices is
supposed.

For instance, if $V=A,$ using the pairing  \eqref{A-dual}, the last
expression reduce to \beqa \nonumber ([A ,W], {\cal F} {\cal B}) &=&
\delta_{m1} \delta_{n0} \sum_{w,w^\prime} \Gamma^{k,00}_{0w; 0
w^\prime} \ \biggl[ \Lambda^r_{0,0,{\tilde m} ; k^\prime, l^\prime,
m^\prime, j^\prime} ( W, b^{ k^\prime} a^{l^\prime + s} d^{m^\prime
+ t} c^{j^\prime + u}) \\  &-&
 \Lambda^r_{{\tilde
k},{\tilde l}, {\tilde m}; 0, l^\prime, m^\prime, 0}\  { s \choose
l^\prime + s } \ (W, b^{\tilde k} a^{{\tilde l} + l^\prime + s}
d^{{\tilde m} + t} c^{u})\biggr]. \label{AWFB} \eeqa Then,  using
\eqref{tildeA-dual}, we get \beqa \nonumber ([A ,{\tilde A}], {\cal
F} {\cal B}) &=& \delta_{m1} \delta_{n0}
\delta_{u0} \sum_{w,w^\prime} \Gamma^{k,00}_{0w; 0 w^\prime}  \\
\nonumber &\times& \sum_{{\tilde m},m^\prime, {\tilde l}, l^\prime}
\biggl[ \Lambda^r_{0,0,{\tilde m} ; 0, l^\prime, m^\prime,0}
\delta_{({l^\prime} + s), 1} -
 \Lambda^r_{0,{\tilde l}, {\tilde m}; 0, l^\prime, m^\prime, 0}\  { s \choose
l^\prime + s } \delta_{({\tilde l} +  l^\prime +s),1} \biggr] \\
&=& \delta_{m1} \delta_{n0} \delta_{u0} \sum_{w,w^\prime}
\Gamma^{k,00}_{0w; 0 w^\prime} \ \sum_{{\tilde m},m^\prime} \biggl[
\Lambda^r_{0,0,{\tilde m} ; 0, 1, m^\prime,0} -
 \Lambda^r_{0,1, {\tilde m}; 0, 0, m^\prime, 0} \biggr] =0, \qquad \eeqa
where the last step follow from the fact \cite{HLR96} \be
\sum_{{\tilde m},m^\prime} \ \Lambda^r_{0,0,{\tilde m} ; 0, 1,
m^\prime,0}  = \sum_{{\tilde m},m^\prime}\ \Lambda^r_{0,1, {\tilde
m}; 0, 0, m^\prime, 0} =0. \ee  Moreover, from \eqref{AWFB} and
(\ref{tildeB-dual}-\ref{tildeC-dual}), we get \be  ([A ,{\tilde B}],
{\cal F} {\cal B}) = \delta_{m1} \delta_{n0} \delta_{s0} \delta_{u0}
\sum_{w,w^\prime} \Gamma^{k,00}_{0w; 0 w^\prime} \sum_{{\tilde
m},m^\prime} \biggl[ \Lambda^r_{0,0,{\tilde m} ;1, 0, m^\prime,0} -
\Lambda^r_{1, 0, {\tilde m}; 0, 0, m^\prime, 0} \biggr]=0, \ee since
\be \sum_{{\tilde m},m^\prime} \ \Lambda^r_{0,0,{\tilde m} ;1, 0,
m^\prime,0}  =
 \sum_{{\tilde m},m^\prime} \ \Lambda^r_{1, 0, {\tilde m}; 0, 0, m^\prime, 0}= \delta_{r 1} ,\ee
and \beqa \nonumber  ([A ,{\tilde C}], {\cal F} {\cal B}) &=&
\delta_{m1} \delta_{n0} \delta_{s0} \sum_{w,w^\prime}
\Gamma^{k,00}_{0w; 0 w^\prime} \sum_{{\tilde m},m^\prime} \biggl[
\Lambda^r_{0,0,{\tilde m} ;0, 0, 0, m^\prime,j^\prime}
\delta_{(j^\prime + u),1} - \Lambda^r_{0, 0, {\tilde m}; 0, 0,
m^\prime, 0} \delta_{u1}
  \biggr]\\ &=& \delta_{m1} \delta_{n0} \delta_{s0} \delta_{u0} \ \sum_{w,w^\prime}
\Gamma^{k,00}_{0w; 0 w^\prime} \sum_{{\tilde m},m^\prime}
\Lambda^r_{0,0,{\tilde m} ; 0, 0, m^\prime,1}=0 , \eeqa since \be
\sum_{{\tilde m},m^\prime} \ \Lambda^r_{0,0,{\tilde m} ; 0, 0,
m^\prime,1}  =0. \ee Finally, from \eqref{AWFB} and
\eqref{tildeD-dual}, we also get \be \nonumber ([A ,{\tilde D}],
{\cal F} {\cal B}) = \delta_{m1} \delta_{n0} \delta_{s0}\delta_{u0}
\sum_{w,w^\prime} \Gamma^{k,00}_{0w; 0 w^\prime} \sum_{{\tilde
m},m^\prime}   \biggl[ {\tilde m} \ \Lambda^r_{0,0,{\tilde m} ;0, 0,
m^\prime,0} - m^\prime \ \Lambda^r_{0, 0, {\tilde m}; 0, 0,
m^\prime, 0}
  \biggr]=0, \ee
where now the last step follow from the fact \be \sum_{{\tilde
m},m^\prime}  \  {\tilde m} \ \Lambda^r_{0,0,{\tilde m} ;0, 0,
m^\prime,0} = \sum_{{\tilde m},m^\prime}  \
 m^\prime \
\Lambda^r_{0, 0, {\tilde m}; 0, 0, m^\prime, 0} =0. \ee

In the same way,  we can show that $ ([C ,{\tilde W}], {\cal F}
{\cal B})=0, \ \forall \ W \in \{{\tilde A},{\tilde B},{\tilde
C},{\tilde D}\} $ and, without greater difficulties, that $([V
,{\tilde W}], {\cal F} {\cal B})=0, \ \forall \ V \in \{B, D\}
 \ {\rm and} \ \forall  W \  \in \{{\tilde A},{\tilde
B},{\tilde C},{\tilde D}\}. $

From the above results, we observe  that the direct sum structure of
the initial non deformed quantum Lie superalgebra  is preserved in
the deformation process. Thus, using directly the results of
reference \cite{HLR96}, we get the non zero super-commutation
relations \beqa \nonumber \{A , C \} &=& {\sinh(2 B \sqrt{xz}) \over
2 \sqrt{xz}},
\\ \nonumber A^2 &=& {1 - \cosh( 2B \sqrt{xz}) \over
4 z}, \qquad C^2 = {1 - \cosh( 2B \sqrt{xz}) \over 4 x}, \\
\label{def-fer-rel} \ds [H, A] &=&  - {1\over 2} A  (1 + \cosh( 2B
\sqrt{xz})) - x {\sinh( 2 B \sqrt{xz})  \over 2 \sqrt{xz}} C, \\
\nonumber \ds
 [H , C] &=&  {1\over 2} C  (1 + \cosh( 2B \sqrt{xz}))   +  z  {\sinh( 2
B \sqrt{xz})  \over 2 \sqrt{xz}} A, \eeqa \beqa [{\tilde A},
{\tilde C} ] &=& {e^{{(p+q)} {\tilde B}} - 1 \over p+q}, \nonumber
\\ \ds [ {\tilde H} , {\tilde A}] &=& - {\tilde A} +  \rho { e^{p
{\tilde B}} \over p+q} \left( {e^{q {\tilde B}}-1 \over q} +
{e^{-p {\tilde B}}-1 \over p} \right), \label{def-bos-rel}
\\ \ds [{\tilde H},{\tilde C} ] &=&  {\tilde C} +  \tau { e^{q
{\tilde B}} \over p+q} \left( {e^{p {\tilde B}}-1 \over p} + {e^{-
q {\tilde B}}-1 \over q} \right). \nonumber \eeqa

\section{Novel Fock  space representation of the type I-II deformed
quantum superalgebra }\label{sec-Fock} The deformed quantum
superalgebra (\ref{def-fer-rel}-\ref{def-bos-rel}) can be
represented in the Fock superspace spanned by the set of states $\{|
n ; j \rangle\}_{n=0}^{\infty}, j=0,1$, in terms of the usual
fermionic ($b,b^\dagger$) and bosonic ($a,a^\dagger$) annihilation
and creation operators associated to the supersymmetric and standard
harmonic oscillator systems, respectively. Let us recall that the
action of these operators  on the Fock superspace is given by \be b
\ |n ; 0 \rangle =0, \qquad b \ |n,1\rangle =|n , 0 \rangle, \qquad
b^\dagger \ |n , 1 \rangle
 =0, \qquad  b^\dagger \ |n ; 0 \rangle = |n ; 1 \rangle \ee and  \be
a \ |n , j \rangle = \sqrt{n} \ | n-1 , j\rangle, \qquad a^\dagger \
|n , j \rangle =\sqrt{n+1} \ | n+1 , j\rangle. \ee

Thus the fermionic oscillator sub-superalgebra can be realized in
terms of the usual fermionic operators $b, b^\dagger$ and  the
identity $I,$ by taking, for instance (see Appendix \ref{sec-appb}),
\beqa \label{A-def-1} A &=& b +  {1 - \cosh(2 \sqrt{xz}) \over 4 z}
\ b^\dagger, \qquad B =I, \qquad  H= \cosh(\sqrt{xz}) \ b^\dagger b,
\\ C &=&    {\cosh( \sqrt{xz}) + 1 \over 2\sqrt{xz}} \sinh(\sqrt{xz})  \
b^\dagger - \sqrt{\frac{z}{x}} \  {\sinh(\sqrt{xz}) \over
\cosh(\sqrt{xz})+ 1} \  b . \label{C-def-1} \eeqa When $x \ne 0,$
another choice for $A $ and $C$  (see Appendix \ref{sec-appb}) is
\be \label{A-def-2} A = {\sinh(\sqrt{x z}) \over \sqrt{2 z}} (b -
b^\dagger ), \qquad C = {\cosh(\sqrt{xz}) + 1 \over \sqrt{2x}} \
b^\dagger - {\cosh(\sqrt{xz}) - 1 \over \sqrt{2x}} \ b.   \ee Let us
insist on the fact that   we have  chosen two realizations for which
$A$ and $C$ are linear combinations of $b$ and $b^\dagger.$ In the
first cas (\ref{A-def-1}-- \ref{C-def-1}), $A$ is a deformation of
$b$ and $C$ of $b^\dagger ,$ whereas in the second one
\eqref{A-def-2}, when $z$ goes to zero, $A \mapsto \sqrt{x \over 2}
\ (b-b^\dagger) $ and $C \mapsto \sqrt{2 \over x } \ b^\dagger.$

On the other hand, a natural realization of the bosonic oscillator
subalgebra in terms of the operators $a, a^\dagger$ and $I,$ is
given by \beqa \label{Atil-def} {\tilde A} &=& \sqrt{\omega}\ a,
\qquad {\tilde B} = I, \qquad
 {\tilde C} =  \sqrt{\omega} \ a^\dagger,
 \qquad {\tilde H} = a^\dagger a + {\tilde \tau} \ a - {\tilde \rho} \
a^\dagger, \eeqa where $ w= {e^{(p+q)} -1 \over p+q} $ and \be
{\tilde \tau} = {\tau \over \sqrt{\omega}} { e^{q} \over p+q} \left(
{e^{p}-1 \over p} + {e^{- q} - 1 \over q} \right), \qquad {\tilde
\rho} = {\rho \over \sqrt{\omega}}{ e^{p} \over p+q} \left( {e^{q}-1
\over q} + {e^{- p} - 1 \over p} \right). \ee Thus, the commutation
relations \eqref{def-bos-rel}, can be obtained from \be [{\tilde A},
{\tilde C}] =\omega I, \qquad [{\tilde H}_0, {\tilde A}] = - {\tilde
A}, \qquad [{\tilde H}_0 , {\tilde C} ] = {\tilde C}, \ee where \be
{\tilde H}_0 = a^\dagger a + {\tilde \tau } \ {\tilde \rho } \ I,
\ee by performing the transformation \be \label{T-trans} T =
\exp\left({\tilde \tau} \ a\right) \ \exp\left({\tilde \rho} \
a^\dagger\right). \ee We notice that $T$ is not an unitary operator
except for  the special case when ${\tilde \rho} = - {\tilde \tau}.$

Let us remark  that these results  represent new realizations of the
deformed fermionic type I (see Appendix \ref{sec-appb}) and deformed
type II quantum Lie algebras that have not been obtained before in
the literature.

It is clear that, in the limit when the deformation parameters goes
to zero, in all the cases, we regain essentially the  standard
representation of the super-oscillator Lie superalgebra $
ho(2|2,{\mathbb R})\oplus ho(4,{\mathbb R})$: \beqa
\label{repre-fer} A&=&b, \qquad
B=I, \qquad C= b^\dagger, \qquad H = b^\dagger b, \\
\label{repre-bos}
 {\tilde A}&=& a, \qquad  {\tilde B}=I, \qquad {\tilde C}= a^\dagger, \qquad {\tilde H} = a^\dagger a.\eeqa

\setcounter{equation}{0}\section{Deformed coherent states}
\label{sec-cinco}

 To be able to construct super-coherent states
associated to the deformed superalgebra
(\ref{def-fer-rel}--\ref{def-bos-rel}), we construct a suitable
deformed annihilation operator ${\mathbb A}_0 $. A possible choice
is \be {\mathbb A}_0 = {\tilde A} +  A, \ee with $A$ and $\tilde A $
as given by equations \eqref{A-def-1} and \eqref{Atil-def} or
\eqref{A-def-2} and \eqref{Atil-def}, respectively. The first case
corresponds to the deformed super-annihilator associated to the
super-symmetric harmonic oscillator, introduced by Aragone and
Zypmann\cite{kn:ArZy}. The second one corresponds to a deformed
annihilator associated to a generalized Hamiltonian isospectral to
the harmonic oscillator one  but two times degenerated
\cite{kn:NaVh3}.

\subsection{Deformed super-coherent states}
Using \eqref{A-def-1} and \eqref{Atil-def}, the annihilator
${\mathbb A}_0 $ becomes \be {\mathbb A}_0 = \sqrt{\omega}\ a +  b
+ {1 - \cosh(2 \sqrt{xz}) \over 4 z} \ b^\dagger. \ee Thus, the
deformed coherent states $|Z ; x , z ,\omega  \rangle,$ i.e., the
eigenstates of ${\mathbb A}_0 $ associated to the eigenvalue $Z
\in {\mathbb C},$ verify the eigenvalue equation \be \left[
\sqrt{\omega} \ a +  b  +  {1 - \cosh(2 \sqrt{xz}) \over 4 z}  \
b^\dagger \right] \ |Z ; x , z , \omega \rangle = Z \ |Z ; x , z,
\omega \rangle. \label{eigen-Z-def} \ee The two independent
solutions of this eigenvalue equation has been obtained in reference \cite{kn:NaVh3}. They   are given by
the normalized eigenstates \be \label{def-ArZy} \ds |Z ; x , z ,
\omega ; \mp \rangle = D \left( {Z \over \sqrt{\omega}} + {i \over
\sqrt{2 z \omega}} \sinh(\sqrt{xz})\right)  {\left[ |0; 0 \rangle
\mp  {i \over \sqrt{2z}} \sinh(\sqrt{xz}) |0 ; 1 \rangle\right]
\over \sqrt{ 1 + { \sinh^2 (\sqrt{xz}) \over 2 z} } }, \ee   where
$D(\alpha)$ is the usual displacement unitary operator
\cite{kn:pere} \be D(\alpha) = \exp\left(\alpha a^\dagger - {\bar
\alpha} a \right). \ee In  absence of  deformation, i.e., in the
limit when $z$ and $x$ tend to zero, the deformed coherent states
\eqref{def-ArZy} becomes the usual coherent states associated to
the standard harmonic oscillator \be |Z ; 0 , 0 , \omega ; \mp
\rangle =  D \left( {Z \over \sqrt{\omega}}\right) | 0; 0 \rangle.
\ee In this case, there is another  independent solution to the
eigenvalue equation \eqref{eigen-Z-def}. It is obtained by solving
directly $[\sqrt{\omega}  a +  b] |Z\rangle = Z |Z\rangle $ and is
exactly   the super-coherent states, given by Aragone and Zypmann
\cite{kn:ArZy}, associated to the super-symmetric harmonic
oscillator: \be |Z ; 0 , 0 , \omega \rangle =  D \left( {Z \over
\sqrt{\omega}}\right) {\left[| 1 ; 0 \rangle -
|0;1\rangle \right] \over \sqrt{2}}. \ee

\subsection{Isospectral harmonic oscillator systems}
When $x \ne 0, $ using \eqref{A-def-2} and \eqref{Atil-def}, the
annihilator ${\mathbb A}_0 $ becomes \be {\mathbb A}_0 =
\sqrt{\omega}\ a +  {\sinh(\sqrt{x z}) \over \sqrt{2 z}} (b -
b^\dagger ) \ee and verifies the canonical commutation relation
\be \label{cano} [{\mathbb A}_0, {\mathbb A}_0^\dagger ]= w I.\ee
In this case, the deformed coherent states $ |Z ; x , z , \omega
\rangle,$ satisfy the eigenvalue equation \be \left[ \sqrt{\omega}
\ a + {\sinh(\sqrt{x z}) \over \sqrt{2 z}} (b - b^\dagger )\right]
\ |Z ; x , z , \omega \rangle = Z \ |Z ; x , z, \omega \rangle.
\label{eigen-Z-iso} \ee Two independent and orthogonal solutions
to this eigenvalue equation are (see \cite{kn:NaVh3})
\be \label{def-coh-sta} |Z ; x, z , \omega ; \mp \rangle = D
\left( {Z \over \sqrt{\omega}} + i {\sinh(\sqrt{x z}) \over
\sqrt{2 z \omega}} \right)  {\left[ |0; 0 \rangle \mp  i |0 ; 1
\rangle\right] \over \sqrt{2} }. \ee

Let us define the Hermitian Hamiltonian \be {\mathbb H}_0 =
{\mathbb A}_0^\dagger {\mathbb A}_0 = \omega a^\dagger a +
{\sinh^2 (\sqrt{xz}) \over 2z} I +  \sqrt{\omega \over 2z}
\sinh{\sqrt{xz}} (a^\dagger b + a b^\dagger - a^\dagger b^\dagger
- a b).  \ee Then, from \eqref{cano}, we get \be \label{hw-com}
[{\mathbb H}_0 , {\mathbb A}_0 ] = - w {\mathbb A}_0, \qquad
[{\mathbb H}_0 , {\mathbb A}_0^\dagger] = w {\mathbb A}_0^\dagger,
\ee i.e., $ {\mathbb A}_0$ and ${\mathbb A}_0^\dagger$ are the
annihilation and creation operators associated to ${\mathbb H}_0
.$ By construction, the eigenstates of ${\mathbb H}_0 ,$
corresponding to the energy eigenvalue $ E_0  = 0, $ are given by
the deformed coherent states \eqref{def-coh-sta}, when $Z=0,$
i.e.,  \be |0 ; x, z , \omega ; \mp \rangle = D \left(  i
{\sinh(\sqrt{x z}) \over \sqrt{2 z \omega}} \right)  {\left[ |0; 0
\rangle \mp  i |0 ; 1 \rangle\right] \over \sqrt{2} }. \ee Then,
from \eqref{hw-com}, the eigenstates of ${\mathbb H}_0 $
corresponding to the energy eigenvalues $E_n = n w,
n=0,1,2,\ldots,$ are given by \be |E_n ; x, z , \omega ; \mp
\rangle = {{\biggr( {\mathbb A}_0^\dagger \biggl)}^{n} \over
\sqrt{n!}} \ |0 ; x, z , \omega ; \mp \rangle. \ee Thus, ${\mathbb
H}_0 $ is  isospectral to the standard harmonic oscillator
Hamitonian but two times degenerated. We notice that this
Hamiltonian is an element of the $osp(2/2) \sdir sh(2/2)$
superalgebra.

By analogy with the standard harmonic oscillator system, the
coherent states associated to ${\mathbb H}_0 $  can  be
obtained by acting with the unitary displacement operator \be
{\mathbb D} (\alpha) = \exp\left(\alpha {\mathbb A}_0^\dagger -
{\bar \alpha} {\mathbb A}_0 \right)\ee on the zero energy
eigenstate, that is  \be \widetilde{|Z ; x, z , \omega ; \mp
\rangle} = {\mathbb D} \left( {Z\over \sqrt{w}}\right) |0 ; x, z ,
\omega ; \mp \rangle. \ee These last states can also be obtained
from \eqref{def-coh-sta} by acting with the unitary operator \be
{\mathbb U} = \exp\left( {\sqrt{2} \over \omega } \ {\rm Re}\, Z \
{\sinh{\sqrt{xz}} \over \sqrt{z}} (b - b^\dagger ) \right). \ee

Let us now consider the transformation \eqref{T-trans} and define
the Hamiltonian \be {\mathbb H} = {\tilde {\mathbb A}} {\mathbb A}
= T {\mathbb H}_0 T^{-1}, \ee with \be {\mathbb A} = T {\mathbb
A}_0 T^{-1} \quad {\rm and} \quad {\tilde {\mathbb A}} = T
{\mathbb A}_0^\dagger T^{-1} .\ee  This Hamiltonian verifies the
$\eta$--pseudo-Hermiticity property \cite{kn:AMostafazadeh} \be
{\mathbb H}^{\dagger} = \eta {\mathbb H} \eta^{-1}, \ee where
$\eta$ is the Hermitian operator \be \eta = {(T^{-1})}^{\dagger}
T^{-1} = e^{- {\tilde \tau} a^\dagger} e^{- {\tilde \rho} a} e^{-
{\tilde \rho} a^\dagger } e^{- {\tilde \tau} a} . \ee

By applying the transformation $T$ to the commutation relations
\eqref{cano} and \eqref{hw-com}, we get \be \label{hw-com-def}
[{\mathbb A}, {\tilde {\mathbb A}}]= w I, \qquad [{\mathbb H} ,
{\mathbb A}] = - w {\mathbb A}, \qquad [{\mathbb H} , {\tilde
{\mathbb A}}] = w {\tilde {\mathbb A}}, \ee i.e., ${\mathbb A},$
${\tilde {\mathbb A}}$, $I$ and $\mathbb H$ verify the commutation
relation of the oscillator Lie algebra $ho(4, {\mathbb R}).$ Thus,
${\mathbb A}$ and $\tilde {\mathbb A} $ are the annihilation and
creation operators associated to ${\mathbb H}.$ This implies that
the Hamiltonian ${\mathbb H}$ has the same energy spectrum than
${\mathbb H}_0$ and their eigenstates are given by $ T |E_n ; x, z ,
\omega ; \mp \rangle. $ The coherent states associated to ${\mathbb
H}$ are the eigenstates of  ${\mathbb A},$ i.e., $T |Z ; x, z ,
\omega ; \mp \rangle. $ These coherent states include both the $x,z$
and the $\rho, \tau $ deformation parameters  and can also be
considered as a class of coherent states associated to the type I-II
fermionic-bosonic  quantum Lie algebra
(\ref{def-fer-rel}-\ref{def-bos-rel}) in the chosen realization.

\setcounter{equation}{0}\section{Conclusion} In this work, using the
$R$--matrix method, we have computed the two oscillator
fermionic-bosonic type I-II quantum bialgebras  and the
corresponding dual quantum Lie superalgebras. We have shown that, in
general, though the quantum bialgebras associated with a two
independent oscillator group are not the structure of a direct sum
of bialgebras associated to the single oscillator group the dual
quantum Lie algebras have the structure of a direct sum of quantum
Lie algebras associated to the single oscillator group. Then we have
given some examples of the type I-II fermionic-bosonic quantum Lie
algebras and several realizations of them in terms of the usual
fermionic and bosonic creation and annihilation operators. Based on
this realization, we have also found a class of coherent states
associated with the two oscillator quantum group. The other types of
fermionic-bosonic quantum algebras, that we have mentioned in
section \ref{sec-tres}, can be treated in the same way.
\setcounter{equation}{0}\section*{Acknowledgments} The author would
like to thank V\'eronique Hussin for valuable discussions and
suggestions. The author's research was partially supported by
research grants from NSERC of Canada. \nolinebreak

\appendix
\setcounter{equation}{0}\section{Determination of the possible $R$
matrices and the bialgebras structures} \label{sec-appa} From
\eqref{reldef}, setting to zero every linear relation between the
generators, we get the block matrices $R^{i,j}$: \be R^{11} =
\pmatrix{ r^{11}_{11} & 0 & r^{11}_{13}
 & 0 & 0 \cr 0 &  r^{11}_{11} & 0 & 0 & 0 \cr
 0 & 0 &  r^{11}_{11} & 0 & 0 \cr
  0 & 0 & 0 &  r^{11}_{11} & 0 \cr
   0 & 0 &  r^{11}_{53} & 0 &  r^{11}_{11} \cr},
   \qquad
 R^{12} = \pmatrix{ 0 & r^{12}_{12} & r^{12}_{13}
 & 0 & 0 \cr 0 & r^{12}_{22} & r^{12}_{23}
 & 0 & 0   \cr
 0 & 0 &  0 & 0 & 0 \cr
  0 & 0 & r^{12}_{43} &  r^{12}_{44} & 0 \cr
   0 & 0 &  r^{12}_{53} &  r^{12}_{54} & 0 \cr},
 \ee
\be R^{13} = \pmatrix{  r^{13}_{11} & r^{13}_{12} & r^{13}_{13} &
 0 & 0 \cr 0 & r^{13}_{22} & r^{13}_{23}
 & 0  & 0 \cr
 0 & 0 &  r^{13}_{33} & 0 & 0 \cr
  0 & 0 & r^{13}_{43} &  r^{13}_{44} & 0 \cr
   r^{13}_{51}  & 0 &  r^{13}_{53} &  r^{13}_{54} & r^{13}_{55}
   \cr},
\ee \be  R^{14} = \pmatrix{ 0 & 0 &  0 & 0 & 0 \cr 0 & 0 & 0 & 0 &
0 \cr
 0 & 0 & 0 & 0 & 0 \cr
  0 & 0 & r^{14}_{43}& 0 & 0 \cr
   0 & 0 & 0 & 0 & 0 \cr},
\qquad  R^{15} = \pmatrix{ 0 & 0 &  0 & 0 & 0 \cr 0 & 0 & 0 & 0 &
0 \cr
 0 & 0 & 0 & 0 & 0 \cr
 0 & 0 & 0 & 0 & 0 \cr
  0 & 0 & r^{15}_{53}& 0 & 0 \cr
  }, \ee

\be  r^{14}_{43}= r^{15}_{53}, \ee

\be R^{22} = \pmatrix{ r^{22}_{11} & r^{22}_{12} &  r^{22}_{13} &
0 & 0 \cr 0 & r^{22}_{22} & r^{22}_{23} & 0 & 0 \cr
 0 & 0 &  r^{22}_{11} & 0 & 0 \cr
 0 & 0 &  r^{22}_{43}& r^{22}_{44} & 0 \cr
  0 & 0 & r^{22}_{53}& r^{22}_{54} & r^{22}_{11} \cr
  } \qquad
 R^{23} = \pmatrix{ 0 & r^{23}_{12} &  r^{23}_{13} & 0
& 0 \cr 0 & r^{23}_{22} & r^{23}_{23} & 0 & 0 \cr
 0 & 0 &  0 & 0 & 0 \cr
 0 & 0 &  r^{23}_{43}& r^{23}_{44} & 0 \cr
  0 &  r^{23}_{52} & r^{23}_{53}& r^{23}_{54} & 0 \cr
  },
  \ee
\be  R^{21} =  R^{24} = R^{25} = {\boldmath 0}, \qquad
r^{23}_{52}= r^{13}_{51}, \qquad r^{22}_{11}= r^{11}_{11}, \ee

\be R^{33} = \pmatrix{ r^{33}_{11}  & 0 & r^{33}_{13} & 0 & 0 \cr
0 & r^{33}_{11} &  0 & 0 & 0 \cr
 0 & 0 & r^{33}_{11} & 0 & 0 \cr
  0 & 0 & 0 & r^{33}_{11}& 0 \cr
   0 & 0 & r^{33}_{53} & 0 &  r^{33}_{11} \cr},
   \ee
\be R^{31}=  R^{32} =  R^{34} = R^{35} = {\boldmath 0}, \qquad
r^{33}_{11}= r^{11}_{11},  \ee

\be R^{43} = \pmatrix{ 0 & r^{43}_{12} &  r^{43}_{13} &
r^{43}_{14} & 0 \cr 0 & r^{43}_{22} & r^{43}_{23} & 0 & 0 \cr
 0 & 0 &  0 & 0 & 0 \cr
 0 & 0 &  r^{43}_{43}& r^{43}_{44} & 0 \cr
  0 &  0 & r^{43}_{53}& r^{43}_{54} & 0 \cr
  }, \qquad
R^{44} = \pmatrix{ r^{44}_{11} & r^{44}_{12} &  r^{44}_{13} & 0 &
0 \cr 0 & r^{44}_{22} & r^{44}_{23} & 0 & 0 \cr
 0 & 0 &  r^{44}_{11} & 0 & 0 \cr
 0 & 0 &  r^{44}_{43}& r^{44}_{44} & 0 \cr
  0 & 0 & r^{44}_{53}& r^{44}_{54} & r^{44}_{11} \cr
  },
  \ee

  \be  R^{41} =  R^{42} = R^{45} = {\boldmath 0}, \qquad
r^{44}_{11}= r^{11}_{11} \qquad r^{43}_{14}= r^{53}_{15}, \ee

\be
  R^{51} = \pmatrix{ 0 & 0 &   r^{51}_{13} & 0 & 0 \cr 0 & 0
 & 0 & 0 & 0 \cr
 0 & 0 & 0 & 0 & 0 \cr
 0 & 0 & 0 & 0 & 0 \cr
  0 & 0 &0 & 0 & 0 \cr
  } \qquad  R^{52} = \pmatrix{ 0 & 0 &  0 & 0 & 0 \cr 0 & 0 & r^{52}_{23} &
0 & 0 \cr
 0 & 0 & 0 & 0 & 0 \cr
  0 & 0 & 0&  0 & 0 \cr
   0 & 0 & 0 & 0 & 0 \cr},
 \ee
\be R^{53} = \pmatrix{  r^{53}_{11} & r^{53}_{12} & r^{53}_{13} &
 0 &  r^{53}_{15} \cr 0 & r^{53}_{22} & r^{53}_{23}
 & 0  & 0 \cr
 0 & 0 &  r^{53}_{33} & 0 & 0 \cr
  0 & 0 & r^{53}_{43} &  r^{53}_{44} & 0 \cr
   0  & 0 &  r^{53}_{53} &  r^{53}_{54} & r^{53}_{55}  \cr},\ee
\be R^{54} = \pmatrix{ 0 & r^{54}_{12} & r^{54}_{13}
 & 0 & 0 \cr 0 & r^{54}_{22} & r^{54}_{23}
 & 0 & 0   \cr
 0 & 0 &  0 & 0 & 0 \cr
  0 & 0 & r^{54}_{43} &  r^{54}_{44} & 0 \cr
   0 & 0 &  r^{54}_{53} &  r^{54}_{54} & 0 \cr},
\qquad
 R^{55} = \pmatrix{ r^{55}_{11} & 0 & r^{55}_{13}
 & 0 & 0 \cr 0 &  r^{55}_{11} & 0 & 0 & 0 \cr
 0 & 0 &  r^{55}_{11} & 0 & 0 \cr
  0 & 0 & 0 &  r^{55}_{11} & 0 \cr
   0 & 0 &  r^{55}_{53} & 0 &  r^{55}_{11} \cr},
\ee

\be r^{55}_{11}= r^{11}_{11}, \qquad r^{51}_{13}= r^{52}_{23}. \ee

\subsection{Deforming the femionic-bosonic bialgebra}

 We are interested to deform  the Lie superalgebra $ ho(2/2,{\mathbb
 R}) \oplus ho(4,{\mathbb R} ) . $ Choosing as reference point the non deformed
bialgebra (\ref{com-fer}-\ref{com-bos}), we get the following set
of quadratic relations:
%%%%%%%%%%%%%%%%%%%%%%%%%%%%%%%%%%%%%%%%%%%%%%%%%%%%%%%%%%%%%%%%%%%%%%%%%%%%

\beqa \left[ \alpha  , \beta \right] & = & r^{22}_{23} \alpha^2 -
r^{12}_{12} \eta \gamma -
(r^{11}_{13} - r^{12}_{23} - r^{22}_{13} ) \alpha + r^{12}_{13} (1 -\eta),  \\
  \left[ \beta , \gamma \right] &=& - r^{12}_{22} \gamma^2 +
  r^{23}_{23} \eta \alpha -
(r^{12}_{23} + r^{13}_{22} - r^{13}_{33} ) \gamma - r^{13}_{23}
(1- \eta ), \\   \left[ \alpha , \gamma \right] &=& r^{22}_{23}
\eta \alpha - r^{12}_{22} \eta \gamma,
\\     \left[ \eta , \beta \right] &=& r^{22}_{23}  \alpha \eta
-  r^{12}_{22}  \eta \gamma, \\    \left\{ \alpha , \eta \right\}
&=& r^{12}_{22} (\eta - \eta^2 ), \\    \left\{
\gamma , \eta \right\} &=& r^{22}_{23} (\eta - \eta^2 ), \\
  \alpha^2 &=& {1\over 2} r^{12}_{12} (1 -\eta^2) +
r^{12}_{22} \alpha,  \\  \gamma^2 &=& {1\over 2} r^{23}_{23} (1
-\eta^2) + r^{22}_{23} \gamma,\eeqa \beqa  \left[ a , b \right] &
= & r^{44}_{43} a^2 -
(r^{55}_{53} - r^{54}_{43} - r^{44}_{53} ) a + r^{54}_{53} (1 -d),  \\
  \left[ b , c \right] &=& - r^{54}_{44} c^2 -
(r^{54}_{43} + r^{53}_{44} - r^{53}_{33} ) c - r^{53}_{43} (1- d
), \\    \left[ a , c \right] &=& r^{44}_{43} d a - r^{54}_{44} d
c,
\\   \left[ d , b \right] &=& r^{44}_{43}  a d
+  r^{54}_{44}  d c ,\\    \left[ a , d\right] &=& r^{54}_{44} (d
- d^2 ), \\    \left[ c , d \right] &=& r^{44}_{43} (d - d^2 ),
\eeqa
%%%%%%%%%%%%%%%%%%%%%%%%%%%%%%%%%%%%%%%%%%%%%%%%%%%%%%%%%%%%%%%%%%%%%%%
%%%%%%%%%%%%%%%%%%%%%%%%%%%%%%%%%%%%%%%%%%%%%%%%%%%%%%%%%%%%%%%
\beqa
\left[ \eta , a \right] & = & r^{22}_{54} \eta (1-d), \\
\left[ \eta , b \right] & = & r^{22}_{43} a \eta - r^{22}_{54}
\eta c, \\ \left[ \eta , c \right] & = & r^{22}_{43} (d-1) \eta, \\
 \left[ \eta , d \right]&=&0, \\
\left[ \alpha , a \right] &=& r^{22}_{54} \alpha + r^{12}_{44} a +
r^{12}_{54} ( 1 - \eta d ),  \\ \left[ \alpha , c
\right] &=&  r^{12}_{43} (d - \eta ) + r^{22}_{43} d \alpha - r^{12}_{44} \eta c, \\
\left[\alpha, b\right]  &=&   (r^{22}_{53} - r^{11}_{53}) \alpha +
r^{12}_{43} a + r^{22}_{43} a \alpha - r^{12}_{54} \eta c +
r^{12}_{53} (1 - \eta ), \\ \left[ \alpha , d \right] &=&
r^{44}_{12} d (\eta - 1) , \\
 \left[  c , \gamma \right]  &=&   r^{22}_{43} \gamma +  r^{23}_{44} c
+  r^{23}_{43} (  1- d \eta ), \\ \left[ \gamma , a \right] &=&
r^{23}_{44} a \eta - r^{22}_{54} \gamma \eta, \\ \left[ \gamma  ,
b \right] &=& r^{23}_{53}(\eta - 1) - ( r^{22}_{53} + r^{23}_{52}
- r^{33}_{53} ) \gamma - r^{23}_{54} c
 - r^{22}_{54} \gamma c +  r^{23}_{43} a \eta, \\ \left[\gamma , d\right] &=&
r^{44}_{23} (1 - \eta)d, \\  \left[\beta , a\right] &=&
r^{23}_{54} \alpha + ( r^{13}_{44} - r^{13}_{55} ) a  +
r^{13}_{54} ( 1 - d ) +
r^{23}_{44} a \alpha - r^{12}_{54} \gamma d, \\
\left[ \beta , c \right] &=&  -  r^{12}_{43} \gamma + (
r^{13}_{33} - r^{13}_{44} - r^{14}_{43}) c + r^{13}_{43}(d-1) -
r^{12}_{44} \gamma c + r^{23}_{43} d \alpha, \\
\nonumber \left[ \beta , b \right] &=& r^{23}_{53} \alpha - (
r^{13}_{51} + r^{11}_{53} - r^{33}_{53} ) \beta  - r^{12}_{53}
\gamma + r^{13}_{43} a \\ &+& (r^{13}_{33} - r^{15}_{53} -
r^{13}_{55}) b - r^{13}_{54} c  - r^{12}_{54} \gamma c +
r^{23}_{43} a \alpha ,\\ \left[ \beta , d \right] &=& r^{44}_{23}
(1 -\eta) d.
 \eeqa
%%%%%%%%%%%%%%%%%%%%%%%%%%%%%%%%%%%%%%%%%%%%%%%%%%%%%%%%%%%%%%%%%%%%%%%%%%%%%%%%%%%%%%%%%%%%%%%%%%%%%%%%
%%%%%%%%%%%%%%%%%%%%%%%%%%%%%%%%%%%%%%%%%%%%%%%%%%%%%%%%%%%%%%%%%%%%%%%%%%%%%%%%%%%%%%%%%%%%%%%%%%%%%%%%
We have also the consistency relations \beqa \label{fer}
r^{23}_{23} r^{12}_{22}&=&0, \qquad r^{12}_{12} r^{22}_{23}=0, \\
r^{12}_{13} + r^{13}_{12} &=&0, \qquad r^{22}_{12} = r^{12}_{22},
\qquad r^{23}_{22} = r^{22}_{23}, \\ r^{13}_{11} + r^{11}_{13} &=&
r^{13}_{33} +
r^{33}_{13}, \qquad   r^{13}_{23} + r^{23}_{13} = 0, \\
 2 r^{22}_{23} r^{12}_{22} &=& ( r^{13}_{11} - r^{23}_{12} - r^{13}_{22}) + ( r^{11}_{13} - r^{12}_{23} - r^{22}_{13}) ,
\eeqa \beqa \label{bos} r^{44}_{54} &=& - r^{54}_{44}, \qquad
r^{43}_{44} = - r^{44}_{43}, \qquad r^{54}_{54}=r^{43}_{43}=0, \\
 r^{53}_{55} + r^{55}_{53} &=&
r^{53}_{33} + r^{33}_{53},\qquad r^{53}_{43} + r^{43}_{53} = 0,
\qquad r^{54}_{53} + r^{53}_{54}=0, \eeqa \be  r^{53}_{55} -
r^{43}_{54} - r^{53}_{44} =  r^{54}_{43} + r^{44}_{53} -
r^{55}_{53}=p, \qquad r^{54}_{43} + r^{53}_{44} - r^{53}_{33} =
r^{33}_{53} - r^{43}_{54} - r^{44}_{53}=q, \ee \beqa r^{43}_{22} +
r^{22}_{43} =0, \qquad r^{54}_{22} + r^{22}_{54} =0, \\
r^{44}_{23} +
r^{23}_{44} =0, \qquad r^{12}_{44} + r^{44}_{12} =0, \\
r^{43}_{23} + r^{23}_{43} =0, \qquad
 r^{12}_{54} + r^{54}_{12} =0, \\
 r^{23}_{54} + r^{54}_{23} = - r^{44}_{23} r^{22}_{54}, \qquad   r^{12}_{43} + r^{43}_{12} = - r^{22}_{43}
 r^{44}_{12},  \\  r^{12}_{53} + r^{53}_{12} = - r^{12}_{54} r^{43}_{22},\qquad
r^{13}_{54} + r^{54}_{13} = - r^{12}_{54}
  r^{44}_{23}, \\
 r^{13}_{43} + r^{43}_{13} = r^{12}_{44} r^{23}_{43},\qquad  r^{53}_{23} + r^{23}_{53} = - r^{23}_{43}
 r^{54}_{22},
 \eeqa
 \beqa
  r^{13}_{44} + r^{44}_{13} - r^{13}_{55} - r^{55}_{13} = r^{12}_{44}
 r^{23}_{44}, \qquad  r^{13}_{33} + r^{33}_{13} - r^{13}_{44} - r^{44}_{13}- r^{14}_{43} - r^{43}_{14} = - r^{12}_{44}
 r^{23}_{44}, \\   r^{22}_{53} + r^{53}_{22} - r^{11}_{53} - r^{53}_{11} = r^{22}_{43}
 r^{22}_{54}, \qquad  r^{53}_{33} + r^{33}_{53} - r^{53}_{22} - r^{22}_{53}- r^{52}_{23} - r^{23}_{52} = - r^{22}_{43}
 r^{22}_{54}.
 \eeqa

There are several possible choices for the remaining entries of
these $R$ matrices so that they verify the QYBE. Indeed, the
different forms of fixing the coefficients in equations equations
\eqref{fer} and \eqref{bos} determine the different types of
deformation of the fermionic-bosonic bialgebra
(\ref{com-fer}-\ref{com-bos}), i.e., the type I--I, I--II, I--III
II--I II-II and II--III.

\setcounter{equation}{0}\section{Realization of the fermionic type I
quantum superalgebra} \label{sec-appb} In this section we give some
realizations for the type I fermionic quantum superalgebra
\eqref{def-fer-rel}, in terms of the  physical generators
\eqref{repre-fer}.

Let us consider the linear combinations \beqa \label{com-a} A &=&
a_0 I + a_1 b + a_2 b^\dagger +
a_3 b^\dagger b, \\ B &=& I, \\
C &=& c_0 I + c_1 b + c_2 b^\dagger + c_3 b^\dagger b, \\
H &=& h_0 I + h_1 b + h_2 b^\dagger + h_3 b^\dagger b,
\label{com-h} \eeqa where $a_0 , a_1 , a_2 , a_3 ,  c_0 , c_1 ,
c_2 , c_3 , h_0 , h_1 , h_2$ and $ h_3 $ are, in general, complex
coefficients to determine. Inserting  (\ref{com-a}--\ref{com-h})
into \eqref{def-fer-rel}, we get the following system of algebraic
equations \beqa a_3 &=& - 2 a_0,  \qquad c_3 = - 2 c_0, \\   a_0^2
+ a_1 a_2 &=& {1 - \cosh (2 \sqrt{xz}) \over 4 z }, \\ c_0^2 + c_1
c_2
&=& {1 - \cosh (2 \sqrt{xz}) \over 4 x },\\
2 a_0 c_0 + a_1 c_2 + a_2 c_1 &=& {\sinh(2 \sqrt{xz}) \over 2
\sqrt{xz}},\\
h_1 a_2 -h_2 a_1 &=& - {1\over 2} (1 + \cosh(2 \sqrt{xz}) a_0 - x
{\sinh(2 \sqrt{xz}) \over 2 \sqrt{xz}} c_0, \\ h_3 a_1 + 2 h_1 a_0
&=& {1\over 2} (1 + \cosh(2 \sqrt{xz})  a_1 + x {\sinh(2
\sqrt{xz}) \over 2 \sqrt{xz}} c_1, \\  h_3 a_2 + 2 h_2 a_0 &=& -
{1\over 2} (1 + \cosh(2 \sqrt{xz})  a_2 - x {\sinh(2
\sqrt{xz}) \over 2 \sqrt{xz}} c_2, \\
h_1 c_2 - h_2 c_1 &=& \sqrt{ z \over x} {\sinh(2 \sqrt{xz}) \over
2} a_0 +  {1\over 2} (1 + \cosh(2 \sqrt{xz}) c_0 , \\ h_3 c_1 + 2
h_1 c_0 &=&  - \sqrt{ z \over x} {\sinh(2 \sqrt{xz})
\over 2} a_1 -  {1\over 2} (1 + \cosh(2 \sqrt{xz}) c_1 , \\
 \\  h_3 c_2 + 2 h_2 c_0 &=&   \sqrt{ z \over x} {\sinh(2 \sqrt{xz}) \over
2} a_2 +  {1\over 2} (1 + \cosh(2 \sqrt{xz}) c_2. \eeqa

There are several solutions to this system of equations which
depend on a subset of arbitrary coefficients. For example, \beqa
c_1 = c_2 = 0= h_3=0, \qquad a_0 = {\pm i} {\cosh\sqrt{xz} \over
\sqrt{2z}}, \qquad a_1 = {1 \over 2 z a_2}, \\
c_0 = \mp i{\sinh\sqrt{xz} \over \sqrt{2x}}, \qquad h_1 = \mp i
{\cosh\sqrt{xz} \over 2 \sqrt{2z} a_2}, \qquad h_2 = \pm i
\sqrt{z\over 2} a_2 ,  \cosh\sqrt{xz}, \eeqa for arbitrary $h_0$
and $a_2 \ne 0. $ Thus, a realization of the fermionic
superalgebra  \eqref{def-fer-rel}, is \beqa A &=& \pm i
{\cosh\sqrt{xz}\over \sqrt{2z}} \ I  + { 1 \over 2 z a_2 } \ b +
a_2 \ b^\dagger \mp 2 i {\cosh\sqrt{xz}  \over \sqrt{2z}} \
b^\dagger b,
\\
B&=&I,
\\ C&=& \mp i {\sinh\sqrt{xz} \over \sqrt{2x}} \ I  \pm 2 i
{\sinh{\sqrt{xz}} \over \sqrt{2x}} \ b^\dagger b,\\
H &=& h_0 \ I \mp i {\cosh{\sqrt{xz}} \over 2 a_2 \sqrt{2z} } \ b
\pm \sqrt{z\over 2} a_2 \cosh\sqrt{xz} \ b^\dagger. \eeqa

In the case when $a_0 =c_2 =0,$ we have \beqa a_1 &=& {1 -
\cosh(2\sqrt{xz}) \over 4 z a_2},  \qquad c_0 = - i
{\sinh\sqrt{xz} \over \sqrt{2x}}, \qquad c_1 = {\sinh\sqrt{xz}
\over 2 a_2 \sqrt{xz}},\\h_1 &=& i  { \cosh(4\sqrt{xz}) - 1 \over
16 a_2 \sqrt{2z} \cosh\sqrt{xz}}, \qquad h_2 =  i a_2 \sqrt{z\over
x}  {
\cosh(4\sqrt{xz}) - 1 \over 4  \sqrt{2} \cosh\sqrt{xz}}, \\
h_3 &=& - {\cosh(3 \sqrt{xz}) + 3 \cosh\sqrt{xz} \over 4
\cosh\sqrt{xz}},\eeqa for arbitrary $h_0$ and $a_2 \ne 0.$ In this
case, the realization of the deformed Lie algebra
 is given by \beqa A&=& {1 - \cosh(2\sqrt{xz})
\over 4 z a_2} \ b +  a_2 \  b^\dagger,\\
B&=&I, \\
C&=& - i {\sinh\sqrt{xz} \over \sqrt{2x}} \ I +  {\sinh\sqrt{xz}
\over 2 a_2 \sqrt{xz}} \ b + 2 i {\sinh\sqrt{xz} \over \sqrt{2x}}
\ b^\dagger b, \\  H &=& h_0 \ I +  i  { \cosh(4\sqrt{xz}) - 1
\over 16 a_2 \sqrt{2z} \cosh\sqrt{xz}} \ b  +  i a_2 \sqrt{z\over
x}  {\cosh(4\sqrt{xz}) - 1 \over 4  \sqrt{2} \cosh\sqrt{xz}} \
b^\dagger   \nonumber \\ &-& {\cosh(3 \sqrt{xz}) + 3
\cosh\sqrt{xz} \over 4 \cosh\sqrt{xz}} \ b^\dagger b.\eeqa  In the
case when $c_0=a_2=0,$ we have \beqa a_0 &=& \pm i {\sinh\sqrt{x
z} \over \sqrt{2 z}},\qquad  a_1 = - 2 \sqrt{x \over z} \ c_1
\coth\sqrt{x z}, \qquad  c_2 = - {\sinh^2 \sqrt{xz}\over 2x c_1} ,
\\  h_1 &=& \mp i \sqrt{x \over 2} \ c_1 \cosh\sqrt{xz}, \qquad
h_2 = \mp  i {\sinh^2 \sqrt{x z} \over 2 \sqrt{2x}  \ c_1}
\cosh\sqrt{xz}, \\ h_3  &=&  \cosh^2\sqrt{xz}, \eeqa for arbitrary
$h_0$ and $c_1 \ne 0. $ The realizations of
\eqref{def-fer-rel}, are given by \beqa A &=& \pm i {\sinh\sqrt{x
z} \over \sqrt{2 z}} \ I - 2 \sqrt{x \over z} c_1 \coth\sqrt{x z}
\ b \mp  2 i {\sinh\sqrt{x z}
\over \sqrt{2 z}} \ b^\dagger b, \\
B&=&I, \\
C &=& c_1 \ b - {\sinh^2 \sqrt{xz}\over 2x c_1} \ b^\dagger, \\
H &=& h_0 \ I  \mp i \sqrt{x \over 2} \ c_1 \cosh\sqrt{xz} \ b \mp
i {\sinh^2 \sqrt{x z} \over 2 \sqrt{2x}  \ c_1} \cosh\sqrt{xz} \
b^\dagger \nonumber \\ &+& \cosh^2\sqrt{xz} \ b^\dagger b. \eeqa

On the other hand if we take $a_0 = c_0 = 0,$ we get \beqa a_1 &=&
{(\cosh\sqrt{xz} \mp 1 )  \over 2 c_2 \sqrt{xz}} \sinh\sqrt{xz},
\qquad a_2 = -
\sqrt{x\over z} \ c_2 \ {\cosh\sqrt{xz} \pm 1 \over \sinh\sqrt{xz}}, \\
c_1 &=& - {\sinh^2\sqrt{xz} \over 2 x c_2}, \qquad h_1 = h_2=0,
\qquad h_3 =\cosh \sqrt{xz}, \eeqa with arbitrary $h_0$ and $ c_2
\ne 0.$ In this case, the  realizations of de deformed Lie algebra
\eqref{def-fer-rel}, are given by \beqa
 A &=& \label{A-real-3}
{(\cosh\sqrt{xz} \mp 1 ) \over 2 c_2 \sqrt{xz}} \sinh\sqrt{xz} \ b
 - \sqrt{x \over z} \ c_2 \ {\cosh\sqrt{xz} \pm 1 \over \sinh\sqrt{xz}} \
 b^\dagger \\
 B&=&I, \\
 C&=&  - {\sinh^2\sqrt{xz} \over 2 x c_2} \ b  + c_2 \ b^\dagger, \\
 H&=& h_0 \ I + \cosh\sqrt{xz} \ b^\dagger b. \eeqa Choosing the superior sign in  \eqref{A-real-3}, $h_0=0$ and
  \be c_2
 =  { ( \cosh\sqrt{xz} - 1 ) \over 2 \sqrt{xz}} \sinh\sqrt{x
 z}, \ee we get the  realization  (\ref{A-def-1}--\ref{C-def-1}),
 whereas if we chosse \be c_2 = {\cosh\sqrt{xz} + 1 \over \sqrt{2x}}, \ee we get
 the realization  \eqref{A-def-2}.

\end{document}